\documentclass[12pt]{iopart}
\usepackage{color,iopams,epsfig}

\begin{document}

\title{{\bf {An Algebraic Pairing Model with }}$Sp(4)$ {\bf {Symmetry and its
Deformation}}}
\author{K. D. Sviratcheva$^1$, A. I. Georgieva$^{1,2}$ and J. P. Draayer$^1$}
\address{$^1${\it Louisiana State University, Department of Physics and Astronomy,
Baton Rouge, Louisiana, 70803-4001 USA}
\\
$^2${\it Institute of Nuclear Research and Nuclear Energy,
Bulgarian Academy of Sciences, Sofia 1784, Bulgaria}}

\begin{abstract}
A fermion realization of the compact symplectic $sp(4)$ algebra provides a
natural framework for studying isovector pairing correlations in nuclei.
While these correlations manifest themselves most clearly in the binding
energies of $0^+$ ground states, they also have a large effect on the energies
of excited states, including especially excited $0^+$ states. In this
article we consider non-deformed as well
as deformed algebraic descriptions of pairing through the reductions
of $sp_{(q)}(4)$ to different realizations of
$u_{(q)}$(2) for single-$j$ and multi-$j$ orbitals.  The model yields
a classification scheme for
completely paired $0^{+}$ states of
even-even and odd-odd nuclei in the $1d_{3/2}$, $1f_{7/2}$, and
$1f_{5/2}2p_{1/2}2p_{3/2}1g_{9/2}$ shells. Phenomenological non-deformed and deformed
isospin-breaking
Hamiltonians are expressed in terms of the generators of the
dynamical symmetry groups $Sp(4)$
and $Sp_{q}(4)$. These Hamiltonians
are related to the most general microscopic pairing problem,
including isovector pairing and isoscalar proton-neutron interaction along with
non-linear interaction in the deformed extension. 
In both the non-deformed and deformed cases the eigenvalues of the Hamiltonian are fit to
the relevant Coulomb corrected experimental $0^{+}$ energies and this, in turn, allows us to
estimate the interaction strength parameters, to
investigate isovector-pairing properties and symmetries breaking, and to predict the
corresponding energies.  While the non-deformed theory yields
results that are comparable to other theories for light nuclei, the deformed
extension, which takes into account higher-order interactions between the
particles, gives a better fit to the data. The multi-shell
applications of the model provide for reasonable predictions of energies of exotic
nuclei.
\end{abstract}

\maketitle

\section{Introduction}

The pairing problem, which was introduced first in atomic physics
\cite{Racah}, was
later applied to nuclear physics
\cite{Flowers,BohrMottelsonPinesBelyaev} in an attempt to
describe binding energies of nuclei and their low-lying vibrational spectra
\cite{deShalitTalmi,Lane}. Recently, there has been 
renewed interest in this
problem because of new experimental studies of exotic nuclei with relatively
large proton
excess or with $N\approx Z$.  This revival of interest in pairing
follows from the recent development of
radioactive beam facilities and attempts to bridge from nuclear
structure considerations to astrophysical phenomena
\cite{Langanke98,Schatz98}.

Along with approximate mean field solutions
(for a review see \cite{Goodman}), the pairing problem can be
solved exactly by
means of various group theoretical methods, which allow one to
explore the underlying
symmetries.
The $SU(2)$ seniority model
\cite
{Kerman,KermanLawsonMacfarlane,PanDraayerOrmand}
provides for a
good description of nuclei with large
proton or neutron excess, where the like-particle pairing plays a
dominant role. The simple ``quasi-spin"
($SU(2) \sim Sp(2)$) approach \cite{Helmers} not only offers an elegant 
way to understand the
results from the conventional seniority scheme 
\cite{Racah,Flowers,TalmiSimpleModels}, based on
$U(2j+1) \supset Sp(2j+1) \supset SO(3)$, but allows for a 
straightforward expansion to
$SO(5) \sim Sp(4)$ to include protons and neutrons, which otherwise 
has proven to be too
complicated. The generalization to the $SO(5)$ model
\cite{Hecht,Ginocchio,Parikh,Dussel} introduces a relation between
identical-particle and proton-neutron ($pn$) isovector (isospin $\tau =1$)
pairing modes. The addition of an isoscalar ($\tau =0$) $pn$ pairing channel
is described within the framework of the
$SO(8)$ model \cite{Pang,EngelPSVD,PalDJ01}  and the Interacting
Boson Model ($IBM$)
\cite{VanIsackerWarner}.

In the limit of dominant isovector $pn$ pairing correlations, a simple
$SO(5)$ seniority model \cite{EngelLangankeVogel,Dobes} is suitable.
Our goal is to
investigate properties of the isovector pairing interaction
within the context of a
fermion realization of the symplectic $sp(4)$ algebra [which is
isomorphic to $so(5)$].
The model space consists of $J^{\pi }=0^+$ states with pairs
coupled to isospin $\tau =1$;
mixing with $\tau =0$ $pn$ pairs is not included. The importance of
the isovector pairing for
binding energies is suggested by experimental data, namely a $\tau
=1$ ground state for most $N=Z$ odd-odd
nuclei with mass number $A>40$ \cite{ZeldesLirin76,Rudolph,GNarro}, 
and by the results of
various theoretical studies \cite
{CivitareseReboiroVogel,LangankeDKR,PovesMartinezPinedo,MartinezPinedoLV,KanekoHZPRC59,MacchiaveliPLB,MacchiaveliPRC,Vogel}.

While coupling to the isoscalar $pn$ pairing mode may
be important in some cases
\cite{SatulaWyss,SatulaDGMN}, we exclude -- to the best of our
ability -- the ground states
of nuclei that show fingerprints of isoscalar pairing correlations.
In short, the
$Sp(4)$ model is applied to $0^+$ ground states of even-$A$ nuclei
and to the higher-lying
$0^+$ isobaric analog states in most of odd-odd nuclei. We refer to these
states as {\it
isovector-paired} states. In
this regard, it is important to note that the two-body interaction includes an isoscalar
term in addition to the dominant isovector pairing interaction. The
isoscalar $pn$ force is related to
the symmetry energy and is diagonal in the isovector-paired basis
states with good isospin \cite{HasegawaKanekoPRC59}. Diagonal high-$J$
components of the nuclear interaction are also present in the model.
The shell structure and
its dimension play an important role in the construction of the
fermion pairs and their
interaction in accordance with the Pauli principle. The isovector
pairing term is assumed to be particle-hole symmetric, which
enters naturally in points to a decrease in energy with respect to
the mean-field solution of a no-pair theory \cite{HeydeGreiner}.

Limiting cases of $sp(4)$ correspond to different reductions of $sp(4)$
to $u(2)$ and show distinct properties of different coupling modes of
the isovector pairing
interaction; specifically, proton-neutron ($pn$) and like-particle
($pp$ and $nn$) pairing
phenomena.  The theory provides for a classification of nuclear states
with respect to the number valence protons and neutrons
occupying a major shell. The notion of a dynamical
symmetry extends this
picture to include an isospin-breaking phenomenological interaction which is
related to a general microscopic Hamiltonian for the pairing problem.
The final result can be written in terms of the
second order invariants of the subalgebras of $sp(4)$ which then
reduces the problem to an exactly solvable theory.

A $q$-deformation of the classical algebraic structure is introduced. A quantum
extension of the dynamical symmetry approach is realized leading also to an exact
$q$-deformed solution of the problem and its limiting cases. The motivation
behind the $q$-deformed generalization is that, as a  
novel and richer model, it allows us to include non-linear features of the interaction and
to investigate the respective changes this may require in the strength parameters 
and in the pairing gaps. Existing applications of the
$q$-deformed algebraic structures
to the pairing problem \cite {Bonatsos} are restricted mainly to the
$SU_{q}(2)$
limit \cite{Sharma,SharmaSh} of the dynamical symmetry approach presented here
for $Sp_{q}(4)$.

Fully-paired even-even and odd-odd nuclei, $32 \leq A \leq 100$, in $1d_{3/2}$,
$1f_{7/2}$ and
$1f_{5/2}2p_{1/2}2p_{3/2}1g_{9/2}$ orbitals are considered in details in this
investigation. An analysis of the results, obtained by fitting model parameters to
experimental data both in the deformed and non-deformed cases, provides for a reasonable
prediction of the relevant $0^{+}$ state energies of nuclei classified as belonging to
a major shell and gives insight into their pair 
structure and isospin mixing.
It  also estimates the broad limits of applicability of  such a simple 
algebraic model
and its deformed non-linear extension-- which is in agreement with the results of other
theoretical approaches for describing
pairing phenomena in nuclear systems
\cite{TalmiSimpleModels,EngelLangankeVogel,HasegawaKanekoPRC59,TalmiThieberger,SandhuRustgi,Moler}.

The paper is organized as follows. In the next section, the algebraic structure
of the fermion realization of $sp(4)$ and its deformation $sp_{q}(4)$ is
introduced with emphasis on the physical interpretation of the generators. In Section 3 the
application of the algebraic constructions is realized through the introduction of a model
Hamiltonian in both the deformed and non-deformed cases. In Section 4,
the parameters of the Hamiltonians are presented as output of a 
fitting procedure to the
respective experimental energies and the results are analyzed.
A summary of our findings and the main conclusions as well as 
possible further developments of
the approach are discussed in the final section.

\section{Algebraic structure of non-deformed and q-deformed $sp(4)$}

To introduce notation, we start with a brief review of the algebraic
structures that enter into the discussion \cite{fermRealSp4}.  The
$sp(4)$ algebra is realized in terms of the creation (annihilation) fermion
operator $c_{m,\sigma }^{\dagger }$ ($c_{m,\sigma }$) $\sigma =\pm 1,$ where
these operators create (annihilate) a particle of type $\sigma $ in a state of
total angular momentum $j=\frac{1}{2},\frac{3}{2},\frac{5}{2},...,$ with
projection $m$ along the $z$ axis ($-j\leq m\leq j$). They satisfy Fermi
anticommutation relations
\begin{equation}
\begin{array}{ll}
\{c_{m^{\prime },\sigma ^{\prime }},c_{m,\sigma }^{\dagger }\}=\delta
_{m^{\prime },m}\delta _{\sigma ^{\prime },\sigma }\qquad & \{c_{m^{\prime
},\sigma ^{\prime }}^{\dagger },c_{m,\sigma }^{\dagger }\}=\{c_{m^{\prime
},\sigma ^{\prime }},c_{m,\sigma }\}=0
\end{array}
\end{equation}
and Hermitian conjugation is given by $(c_{m,\sigma }^{\dagger
})^{*}=c_{m,\sigma }.$ For a given $\sigma ,$ the dimension of the fermion
space is $2\Omega _{j}=2j+1.$

The deformation of the $sp_{q}(4)$ algebra is introduced in terms of
$q$-deformed creation and annihilation operators $\alpha _{m,\sigma
}^{\dagger }
$ and $\alpha _{m,\sigma },$ $(\alpha _{m,\sigma }^{\dagger })^{*}=\alpha
_{m,\sigma }$, where $\alpha _{m,\sigma }^{(\dagger )}\rightarrow
c_{m,\sigma }^{(\dagger )}$ in the limit $q\rightarrow 1.$ The deformed
single-particle operators are defined through their anticommutation
relation for
every $\sigma $ and $m$ \cite{fermRealSp4}:
\begin{equation}
\begin{array}{ll}
\{\alpha _{m,\sigma },\alpha _{m^{\prime },\sigma }^{\dagger }\}_{q^{\pm
1}}=q^{\pm \frac{N_{\sigma }}{2\Omega _{j}}}\delta _{m,m^{\prime }}\qquad &
\{\alpha _{m,\sigma },\alpha _{m^{\prime },\sigma ^{\prime }}^{\dagger
}\}=0,~\sigma \neq \sigma ^{\prime } \\
\{\alpha _{m,\sigma }^{\dagger },\alpha _{m^{\prime },\sigma ^{\prime
}}^{\dagger }\}=0 & \{\alpha _{m,\sigma },\alpha _{m^{\prime },\sigma
^{\prime }}\}=0,
\end{array}
\label{qfcr}
\end{equation}
where by definition the $q$-anticommutator is given as $\left\{ A,B\right\}
_{q^{k}}=AB+q^{k}BA.$ A property with physics impact is the dependence
of the deformed anticommutation relations on the shell dimension and the
operators that count the number of particles, $N_{\pm
1}=\sum_{m=-j}^{j}c{{{_{m,\pm 1}^{\dagger }}}}c {_{m,\pm 1}}.$
\newline

{\bf Generators of the symplectic group --}  The
generators of $Sp(4)$ and $Sp_{q}(4)$ are
expressed in terms of
non-deformed
\cite{Hecht,KleinMarshalek} and deformed single-particle operators
\cite{fermRealSp4}, respectively,
\begin{equation}
\left.
\begin{array}{l}
A_{\frac{\sigma +\sigma ^{\prime }}{2}}=\frac{1}{\sqrt{2\Omega
_j(1+\delta _{\sigma,\sigma
^{\prime }})}}\sum_{m=-j}^{j}{{(-1)}^{j-m}}c_{m,\sigma }^{\dagger
}c_{-m,\sigma ^{\prime
}}^{\dagger } \\
B_{\frac{\sigma +\sigma ^{\prime }}{2}}=\frac{1}{\sqrt{2\Omega
_j(1+\delta _{\sigma,\sigma
^{\prime }})}}\sum_{m=-j}^{j}{{(-1)}^{j-m}}c{_{-m,\sigma
}}c{_{m,\sigma ^{\prime }}}
\end{array}
\right\}
\begin{array}{c}
non- \\
deformed
\end{array}
\label{g1}
\end{equation}

\begin{equation}
\left.
\begin{array}{l}
F_{\frac{\sigma +\sigma ^{\prime }}{2}}=\frac{1}{\sqrt{2\Omega
_j(1+\delta _{\sigma,\sigma
^{\prime }})}}\sum_{m=-j}^{j}{{(-1)}^{j-m}}\alpha _{m,\sigma }^{\dagger }\alpha
_{-m,\sigma ^{\prime }}^{\dagger } \\
G_{\frac{\sigma +\sigma ^{\prime }}{2}}=\frac{1}{\sqrt{2\Omega
_j(1+\delta _{\sigma,\sigma
^{\prime }})}}\sum_{m=-j}^{j}{{(-1)}^{j-m}}\alpha {_{-m,\sigma
}}\alpha {_{m,\sigma
^{\prime }}}
\end{array}
\right\}
\begin{array}{c}
deformed
\end{array}
\label{qg1}
\end{equation}
where $\sigma ,\ \sigma ^{\prime}=\pm 1$ and $A_{0,\pm 1}=(B_{0,\pm
1})^{\dagger },\ F_{0,\pm
1}=(G_{0,\pm 1})^{\dagger }$. These operators create (annihilate) a
pair of fermions coupled to
total angular momentum and parity $J^{\pi }=0^{+}$ \cite{Hecht,Flowers}  and
thus constitute boson-like objects. The rest of the generators
of $Sp(4)$ are
$
D_{\sigma ,\sigma ^{\prime }}=\frac{1}{\sqrt{2\Omega _j}}
\sum_{m=-j}^{j}c{{{_{m,\sigma
}^{\dagger } }}}c{_{m,\sigma ^{\prime }}},
$
and for $Sp_{q}(4)$ they are
$
E_{\pm 1,\mp 1}=\frac{1}{\sqrt{2\Omega _j}} \sum_{m=-j}^{j}\alpha {{{_{m,\pm
1}^{\dagger }}}}\alpha {_{m,\mp 1}},
$
in addition to the number operators, $N_{\pm 1}$, which remain
non-deformed in this realization of $sp_{q}(4)$.
The ten non-deformed (deformed) generators close on the symplectic $
sp_{(q)}(4)$ algebra with the commutation relations given in \cite
{fermRealSp4}. For nuclear structure applications we use the set
of the $q$-deformed commutation relations that is symmetric with respect to the
exchange of the deformation parameter $q\leftrightarrow q^{-1}.$
\newline

{\bf Physical interpretation of the generators --} When considered to
be a dynamical symmetry, the
$Sp(4)$ symplectic group can be used to describe distinct collective
nuclear phenomena through
different interpretations of the $\sigma$ quantum number in the $Sp(4)\supset
U(1)\otimes SU(2)$ reduction. When $\sigma $ is used to distinguish
between protons ($\sigma =1$)
and neutrons ($\sigma =-1$), the Cartan generators of the $Sp(4)$ group $N_{\pm
1}$ (with eigenvalues $N_{\pm }$) enter as the number of the valence
protons and valence neutrons, respectively.

The significant reduction limits of $sp_{(q)}(4)$ are summarized for
the non-deformed and $q$-deformed cases in Table 1, where by definition
$[X]_{k}=\frac{q^{kX}-q^{-kX}  }{q^{k}-q^{-k}}$ and $\rho _{\pm
}=(q^{\pm 1}+q^{\pm
\frac{1}{2\Omega }})/2$.
Table 1 consists of four different realizations of a two-dimensional
unitary subalgebra
$u_{(q)}^{\mu }(2)\supset u^{\mu }(1)\oplus su_{(q)}^{\mu }(2)$ ($
\mu =\{\tau ,0,\pm \}$) and the corresponding second-order Casimir
invariant of $su_{q}(2)$. The
``classical'' formulae are restored in the limit when $q$ goes to $1$.
In the first realization,
$su^{\tau }(2)$, the
generators ${{\tau }_{0,\pm }}$ are associated with the components of the
isospin of the valence particles. The $SU^{0}(2)$ limit describes
proton and neutron pairs ($pn$), while the $SU^{\pm }(2)$ limit is related to
coupling between identical particles, proton-proton ($pp$) and
neutron-neutron ($nn$) pairs.

\smallskip \noindent \textbf{Table 1.} Realizations of the unitary
subalgebras of $sp_{(q)}(4)$: isospin symmetry $(\mu =\tau )$, $pn$
$(\mu =0)$ and
$pp(nn)$ $(\mu =\pm )$ coupling, along with the Casimir invariants of
$su_{q}^{\mu
}(2)$.
\[
\hspace{-70pt}
\begin{tabular}{|c|c|c|c|c|}
\hline
$\mu $ & $u^{\mu }(1)$ & $su^{\mu }(2)$ & $su_{q}^{\mu }(2)$ & $
\mathbf{C}_{2}(su_{q}^{\mu }(2))\doteq \mathbf{C}_{2,q}^{\mu }$ \\ \hline\hline
$\tau $ & $
\begin{array}{c}
N=\\
N_{+1}{+}N_{-1}
\end{array}
$ & $
\begin{array}{c}
\tau _{\pm }{\equiv }D_{\pm 1,\mp 1} \\
\tau _{0}{=}\frac{N_{1}-N_{-1}}{2}
\end{array}
$ & $
\begin{array}{c}
T_{\pm }{\equiv }E_{\pm 1,\mp 1}  \\
T_{0}\equiv \tau _{0}
\end{array}
$ & $ {\Omega }_{j}(\{T_{+},T_{-}\}{+}\left[ \frac{1}{{\Omega }_{j}}\right]
\left[ T_{0}\right] _{\frac{1}{2{\Omega }_{j}}}^{2})$ \\ \hline
$0$ & $\tau _{0}$ & $
\begin{array}{c}
A_{0},{B}_{0} \\
D_{0}{\equiv }\frac{N}{2}-\Omega _{j}
\end{array}
$ & $
\begin{array}{c}
F_{0},G_{0} \\
K_{0}\equiv D_{0}
\end{array}
$ & $ {\Omega }_{j}(\{G_{0},F_{0}\}{+}\left[ \frac{1}{{\Omega }_{j}}\right]
\left[ K_{0}\right] _{\frac{1}{2{\Omega }_{j}}}^{2}) $ \\ \hline
$\pm $ & $N_{\mp 1}$ & $
\begin{array}{c}
A_{\pm 1},B_{\pm 1} \\
D_{\pm 1}{\equiv }\frac{N_{\pm 1}-\Omega _{j}}{2}
\end{array}
$ & $
\begin{array}{c}
F_{\pm 1},G_{\pm 1} \\
K_{\pm 1}{\equiv }D_{\pm 1}
\end{array}
$ & $ \frac{{\Omega }_{j}}{2}(\{G_{\pm 1},F_{\pm 1}\}{+}\rho _{\pm }\left[
\frac{2}{{\Omega }_{j}}\right] \left[ K_{\pm 1}\right]^2 _{\frac{1}{{\Omega }
_{j}}})$ \\ \hline
\end{tabular}
\]

Within a representation $\Omega _{j}$, the space of
fully-paired states is constructed by the
pair-creation $q$-deformed operators
$F{{{_{0,\pm 1}}}}$ (\ref{qg1}) (non-deformed operators ${A}_{0,\pm
1}$ (\ref{g1})), acting on the
vacuum state
\cite {KleinMarshalek}:
\begin{equation}
\left|n_{1},n_{0},n_{-1}\right) _{q}=\left( F{{{_{1}}}}\right)
^{n_{1}}\left( F{{{_{0}}}}\right) ^{n_{0}}\left( {{{F_{-1}}}}\right)
^{n_{-1}}\left| 0\right\rangle ,  \label{qGencsF}
\end{equation}
where $n_{1},n_{0},n_{-1}$ are the total number of pairs of each kind, $pp$,
$pn$, $nn$, respectively. The basis is obtained by orthonormalization 
of (\ref{qGencsF}). The
$q$-deformed states are in general different from the classical ones 
and coincide with them in
the limit $ q\rightarrow 1.$

The generalization of the pairing problem to multi-shells dimension
\cite{Kerman,TalmiSimpleModels,Ginocchio} leads to a natural expansion of the fermion
realization of the $sp(4)$ algebra, allowing the nucleons to occupy a space of several
orbits. The commutation relations between the ten non-deformed (deformed) generators of
the generalized $Sp_{(q)}(4)$ and the related algebraic formulae (derived in the
single-level realization
\cite {fermRealSp4}) remain the same with the substitution
$\Omega _j \rightarrow \Omega $, where $2\Omega =\sum_{j}2\Omega _{j}=\sum_{j}(2j+1)$.

\section{Theoretical model with $Sp(4)$ dynamical symmetry}

In the deformed and non-deformed cases, the basis states $
\left| n_{1},n_{0},n_{-1}\right) _{(q)}$ (\ref{qGencsF}) give the
isovector-paired
$0^{+}$ states of a nucleus with $N_{+}=2n_{1}+n_{0}$ valence
protons and $N_{-}$ $
=2n_{-1}+n_{0}$ valence neutrons. This yields a simultaneous classification
of the nuclei in a given major shell and of their corresponding isovector-paired
states. The classification scheme is illustrated for the simple cases of $
1d_{3/2}$ with $\Omega _{j=3/2}=2$ (Table 2a) and $1f_{7/2}$ with $\Omega
_{j=7/2}=4$ (Table 2b). The total number of the valence particles,
$ n=N_{+}+N_{-}$, enumerates the rows and the eigenvalue $i$ of the 
third projection of the valence
isospin $\tau_0$ enumerates the columns. Isotopes of an element are situated
along the right diagonals, isotones -- along the left diagonals, and the rows
consist of isobars for a given mass number. The shape of the table is
symmetric with
respect to $i$ (with the exchange $n_{1}\leftrightarrow n_{-1}$), as well as
with respect to $n-2\Omega $ (middle of the shell). This is a consequence
of the charge independent nature of the interaction and the Pauli
principle, respectively.

\smallskip
\noindent
{\bf Table2a.} Classification scheme of nuclei, $\Omega _{3/2}=2.$
\[
\begin{tabular}{|c||cc|c|cc|}
\hline
$n\backslash i$ & 2 & \multicolumn{1}{|c|}{1} & 0 & -1 & \multicolumn{1}{|c|}{-2} \\
\hline\hline
0 &  &  & $
\begin{array}{l}
_{16}^{32}S_{16}
\end{array}
$ &  &  \\ \cline{1-1}\cline{3-5}
2 &  & \multicolumn{1}{|c|}{$
\begin{array}{l}
_{18}^{34}Ar_{16}
\end{array}
$} & $
\begin{array}{l}
_{17}^{34}Cl_{17}
\end{array}
$ & $
\begin{array}{l}
_{16}^{34}S_{18}
\end{array}
$ & \multicolumn{1}{|c|}{} \\ \hline
4 & $
\begin{array}{l}
_{20}^{36}Ca_{16}
\end{array}
$ & \multicolumn{1}{|c|}{$
\begin{array}{l}
_{19}^{36}K_{17}
\end{array}
$} & $
\begin{array}{l}
_{18}^{36}Ar_{18}
\end{array}
$ & $
\begin{array}{l}
_{17}^{36}Cl_{19}
\end{array}
$ & \multicolumn{1}{|c|}{$
\begin{array}{l}
_{16}^{36}S_{20}
\end{array}
$} \\ \hline
6 &  & \multicolumn{1}{|c|}{$
\begin{array}{l}
_{20}^{38}Ca_{18}
\end{array}
$} & $
\begin{array}{l}
_{19}^{38}K_{19}
\end{array}
$ & $
\begin{array}{l}
_{18}^{38}Ar_{20}
\end{array}
$ & \multicolumn{1}{|c|}{} \\ \cline{1-1}\cline{3-5}
8 &  &  & $
\begin{array}{l}
_{20}^{40}Ca_{20}
\end{array}
$ &  &  \\ \hline
\end{tabular}
\]

\vskip .5cm
\noindent
{\bf Table2b.} Classification scheme of nuclei, $\Omega _{7/2}=4$.
The shape of the table is
symmetric with respect to the sign of $i$ and $n-2\Omega $. The basis
states for each nucleus are
labeled by the numbers of particle pairs
$\left| n_{1},n_{0},n_{-1}\right) $.
\[
\begin{tabular}{|c||c|cccc|}
\hline
$n\backslash i$ & 0 & -1 & \multicolumn{1}{|c}{-2} & \multicolumn{1}{|c}{-3} &
\multicolumn{1}{|c|}{-4} \\ \hline\hline
0 & $
\begin{array}{l}
|0,0,0) \\
\begin{array}{l}
_{20}^{40}Ca_{20}
\end{array}
\end{array}
$ &  &  &  &  \\ \cline{1-1}\cline{2-3}\cline{2-3}
2 & $
\begin{array}{l}
|0,1,0) \\
\begin{array}{l}
_{21}^{42}Sc_{21}
\end{array}
\end{array}
$ & $
\begin{array}{l}
|0,0,1) \\
\begin{array}{l}
_{20}^{42}Ca_{22}
\end{array}
\end{array}
$ & \multicolumn{1}{|c}{} &  &  \\ \cline{1-1}\cline{1-1}\cline{1-4}
4 &
\begin{tabular}{l}
$|1,0,1)$ \\
$|0,2,0)$ \\
$
\begin{array}{l}
_{22}^{44}Ti_{22}
\end{array}
$%
\end{tabular}
& $
\begin{array}{l}
|0,1,1) \\
\\
\begin{array}{l}
_{21}^{44}Sc_{23}
\end{array}
\end{array}
$ & \multicolumn{1}{|c}{$
\begin{array}{l}
|0,0,2) \\
\\
\begin{array}{l}
_{20}^{44}Ca_{24}
\end{array}
\end{array}
$} & \multicolumn{1}{|c}{} &  \\ \cline{1-5}
6 &
\begin{tabular}{l}
$|1,1,1)$ \\
$|0,3,0)$ \\
$
\begin{array}{l}
_{23}^{46}V_{23}
\end{array}
$%
\end{tabular}
&
\begin{tabular}{l}
$|1,0,2)$ \\
$|0,2,1)$ \\
$
\begin{array}{l}
_{22}^{46}Ti_{24}
\end{array}
$%
\end{tabular}
& \multicolumn{1}{|c}{$
\begin{array}{l}
|0,1,2) \\
\\
\begin{array}{l}
_{21}^{46}Sc_{25}
\end{array}
\end{array}
$} & \multicolumn{1}{|c|}{$
\begin{array}{l}
|0,0,3) \\
\\
\begin{array}{l}
_{20}^{46}Ca_{26}
\end{array}
\end{array}
$} & \multicolumn{1}{|c|}{} \\ \hline
8 &
\begin{tabular}{l}
$|2,0,2)$ \\
$|1,2,1)$ \\
$|0,4,0)$ \\
$
\begin{array}{l}
_{24}^{48}Cr_{24}
\end{array}
$%
\end{tabular}
&
\begin{tabular}{l}
$|1,1,2)$ \\
$|0,3,1)$ \\
\\
$
\begin{array}{l}
_{23}^{48}V_{25}
\end{array}
$%
\end{tabular}
& \multicolumn{1}{|c}{$
\begin{array}{l}
|0,2,2) \\
|1,0,3) \\
\\
\begin{array}{l}
_{22}^{48}Ti_{26}
\end{array}
\end{array}
$} & \multicolumn{1}{|c|}{$
\begin{array}{l}
|0,1,3) \\
\\
\\
\begin{array}{l}
_{21}^{48}Sc_{27}
\end{array}
\end{array}
$} & \multicolumn{1}{|c|}{$
\begin{array}{l}
|0,0,4) \\
\\
\\
\begin{array}{l}
_{20}^{48}Ca_{28}
\end{array}
\end{array}
$} \\ \cline{1-1}\cline{2-6}
\end{tabular}
\]

\vskip .5cm
{\bf Model Hamiltonian --} As a natural approach within a microscopic
picture, the most general Hamiltonian of a system with $Sp(4)$ symmetry, which
preserves the total
number of particles, can be expressed through the group generators as
following \cite{KleinMarshalek}:
\begin{eqnarray}
H &=&-\epsilon N-GA_{0}B_{0}-F(A_{+1}B_{+1}+A_{-1}B_{-1})-\frac{1}{2}
E(\left\{ \tau _{+},\tau _{-}\right\} -\frac{N}{2\Omega })  \nonumber \\
&&-C\frac{N(N-1)}{2}-D(\tau _{0}\tau _{0}-\frac{N}{4}),  \label{clH}
\end{eqnarray}
where $G,F,E,C$ and $D$ are phenomenological constant interaction strength
parameters ($G\geq 0,F\geq 0$ for attraction), $\epsilon >0$ is a Fermi
level energy.

An important feature of the phenomenological Hamiltonian (\ref{clH}) is that
it not only breaks the isospin symmetry ($D\neq \frac{E}{2\Omega }$) 
but it also mixes states
with definite isospin values ($F\neq G$). This is different from other
applications of non-deformed and deformed $sp(4)$ or $o(5)$ algebras with
isospin-invariant Hamiltonians \cite{Hecht,EngelLangankeVogel,Berej}. 
Although the degree
of mixing is expected to be smaller than for isoscalar-isovector mixing,
it may still add an interesting contribution to the study of
the isospin mixing \cite{Hagberg,OrmandBrown,Lisetskiy}.

Possible applications of the Hamiltonian to real nuclei can be determined
through a detailed investigation of the various terms introduced in
(\ref{clH}).
The first two terms ($G,F$) of the Hamiltonian
(\ref{clH}) account for $J=0$ isovector pairing between
non-identical and identical particles, respectively.
To reflect the assumption that a zero pairing energy corresponds
to a state with no possible
breaking of a pair \cite{HeydeGreiner}, a particle-hole concept is
incorporated in these two terms (but not in the $\epsilon $-, $C$- and $D$-terms).
Hole pair-creation (annihilation) operators can be introduced not only for identical
particle pairs ($pp$ or
$nn$) \cite{HeydeGreiner}, but also for $pn$ pairs.  This corresponds
to a change from the particle
to the hole number operator, $N_{\pm }\rightarrow 2\Omega -N_{\pm }$
for $N_{\pm
}>\Omega $ and $N\rightarrow 4\Omega -N$ for $ N>2\Omega $.

The next term ($E$) can be related to the symmetry energy \cite
{TalmiSimpleModels,Hecht} as its expectation value in states
with definite isospin is
\begin{equation}
\left\langle n,\tau ,i\right| \frac{E}{2}\left\{ \tau _{+},\tau
_{-}\right\} \left| n,\tau ,i\right\rangle =\left\langle E\frac{\tau ^2-\tau
_0^2}{2\Omega }\right\rangle =E\frac{\tau
\left(
\tau +1\right) -i^{2}}{2\Omega },  \label{Eterm}
\end{equation}
which enters as a symmetry term in many nuclear mass relationships
\cite{Janecke,DufloZuker}. The second order Casimir invariant of
$sp(4)$ \cite{GoLiSp} sets linear dependence between the terms in
(\ref{clH}), which
yields to a direct relation between the symmetry and pairing
contributions: a fact that
has been already pointed out in a phenomenological analysis based on
the experimental
nuclear masses and excitation energies
\cite{Vogel}.

The two-body interaction in (\ref{clH}), which is 
written in terms of the
group generators, arises naturally from the microscopic picture. In the
single-$j$ case its form is
\cite{Kerman} 
\begin{eqnarray}
H_{int}^j &=&-\frac{1}{2}\sum_{\left\{ \sigma \right\} }
\left\langle \sigma _{1},\sigma _{2}\left| V\right| \sigma
_{4},\sigma _{3}\right\rangle
\sum_{M,m,m^{\prime }} c_{j,m,\sigma _{1}}^{\dagger }c_{j,M-m,\sigma
_{2}}^{\dagger }c_{j,M-m^{\prime },\sigma _{3}}c_{j,m^{\prime },\sigma _{4}}
\label{Hmicro}
\end{eqnarray}
where $\left\{ \sigma \right\} =\left\{ \left( \sigma _{1},\sigma
_{2},\sigma _{3},\sigma _{4}\right) \right\} =\{{\left( +,+,+,+\right) ,}$ $
\left( +,-,+,-\right) ,$ $\left( +,-,-,+\right) ,$ $\left(
-,-,-,-\right) \}$. The coefficient
$\left\langle \sigma _{1},\sigma _{2}\left| V\right| \sigma
_{4},\sigma _{3}\right\rangle $
is the expectation value of the two-body interaction potential
between pairs of quantum numbers
$\sigma _{4},\ \sigma _{3}$ and $\sigma _{1},\ \sigma _{2}$. The second sum in
(\ref{Hmicro}) can be expanded into three terms. The first term
corresponds to pairing to total angular
momentum $J=0\ (M=0)$, the second term includes high-$J$ ($J\ne 0,\
m=m^{\prime }$) components of
the interaction and can be represented by
$\left\{\tau _{+},\tau _{-}\right\}-N/(2\Omega )$, $N(N-1)$ and $\tau
_{0}\tau _{0}-N/4$. The
rest of the sum is the residual interaction that is neglected. A multi-shell
generalization of the microscopic Hamiltonian
(\ref{Hmicro}) can be related to the phenomenological one (\ref{clH})
and the interaction strengths
can be obtained in terms of the phenomenological parameters
\begin{equation}
\begin{array}{lc}
\begin{array}{c}
J=0\ pairing \\
interaction
\end{array}
& \left\{
\begin{array}{c}
\left\langle ++\left| V_{P_0}\right| ++\right\rangle =\left\langle --\left|
V_{P_0}\right| --\right\rangle =F/\Omega  \\
\left\langle -+\left| V_{P_0}\right| +-\right\rangle =\left\langle -+\left|
V_{P_0}\right| -+\right\rangle =G/\Omega
\end{array}
\right.  \\
&
\begin{array}{l}
\left\langle ++\left| V\right| ++\right\rangle =\left\langle --\left|
V\right| --\right\rangle =C+D/2 \\
\left\langle -+\left| V\right| -+\right\rangle =2C-D,\
\left\langle -+\left| V\right| +-\right\rangle =E/\Omega .
\end{array}
\end{array}
\label{V_FG}
\end{equation}
This connection (\ref{V_FG}) with the interaction matrix elements gives a
real physical meaning to the constant phenomenological strength parameters,
and, therefore, their estimation can lead to a microscopic description of the
nuclear interaction.

The three terms $C,\ D$ and $E$ in (\ref{clH}) that arise from the
dynamical $Sp(4)$ symmetry are
related to the microscopic nature of the $pn$ isoscalar correlations.
As can be clearly seen from the
expression
$\frac{E}{2\Omega }(-{\bf \tau }^2
+\frac{3N}{4}+\frac{1}{2}\frac{N(N-1)}{2})-(C+\frac{E}{4\Omega
})\frac{N(N-1)}{2}-(D-\frac{E}{2\Omega })(\tau ^2_{0}-\frac{N}{4})$
[see (\ref{clH})] and
from relation (\ref{V_FG}), for $D=E/{2\Omega }$ and $C+D/2=0$ we
obtain the $J$-independent $pn$
isoscalar force. It is closely related to
the symmetry energy ($E$), is diagonal in the pairing basis with $G=F$  and
can be compared to
\cite{HasegawaKanekoPRC59,KanekoHasegawaPRC60}. Therefore, the
$Sp(4)$ model interaction consists
of isovector ($pp,\ nn,\ pn$) pairing and isoscalar ($pn$) force in
addition to a possible
isospin-breaking term and
$J>0$ identical-particle pairing correlations.

In this way, the phenomenological Hamiltonian (\ref{clH}) can be used to
describe general properties of the nuclear interaction, which serves as a
motivation to fit the theoretical expectation values of (\ref{clH}) to the
energies of the corresponding $0^{+}$ states of nuclei in a
very broad region.

Within the algebraic framework, the important reduction chains of the
symplectic algebra to the unitary two-dimensional subalgebras allow the
Hamiltonian (\ref{clH}) to be expressed through second-order
operators $C_{2}^{\tau,0,\pm }$ (Table 1):
\begin{eqnarray}
H&=&-\eta _{1}C_{2}^{\tau }-\eta _{2}\tau _{0}^{2}-\eta _{3}C_{2}^{0}-\eta
_{4}D_0^{2}-\eta _{5}(C_{2}^{+}{+}C_{2}^{-})-\eta _{6}(D_{+1}(D_{+1}{-}1)+
\nonumber \\
&&
D_{-1}(D_{-1}{-1}))-\eta _{7}N+\eta _{8}.
\label{CasH}
\end{eqnarray}
The $\eta _{i}$-coefficients ($i=1,2,\ldots ,8$) in (\ref{CasH}) are not
linearly independent; they are related to the phenomenological parameters
of the model (\ref{clH}) in the following way:
\begin{eqnarray}
\eta _{1} &=&\frac{E}{2\Omega };\eta _{2}=(D-\frac{1}{2\Omega }E); \
\eta _{3} =\frac{G}{2\Omega };\eta _{4}=-(\frac{G}{2\Omega }-2C);\
\eta _{5} =-\eta _{6}=\frac{F}{\Omega };  \nonumber \\
\eta _{7} &=&\left\{
\begin{array}{c}
\epsilon -C(1-4\Omega )/2-D/4-(E-G)/(4\Omega ),\
N\leq 2\Omega \\
\epsilon -C(1-4\Omega )/2-D/4+(E-G)/(4\Omega ),\
N>2\Omega ;
\end{array}
\right.  \nonumber \\
\eta _{8} &=&\left\{
\begin{array}{c}
2C\Omega ^{2}+G/2\ \ \ \ \ \ \ \ \ \ \ \ \ \ ,\ N\leq 2\Omega \\
2C\Omega ^{2}+G/2+(E-G),\ N>2\Omega .
\end{array}
\right.  \label{eta_FG}
\end{eqnarray}
The ratios $\eta _{2}/\eta _{1},\eta _{4}/\eta _{3},$ $\eta _{6}/\eta _{5}$
determine the extent to which the symmetry in each limit is broken \cite
{VanIsacker}.

In the $q$-deformed case, a Hamiltonian can be constructed that is analogous to
(\ref{CasH})
and is chosen to coincide with the non-deformed one (\ref{clH}) in the
limit $q\rightarrow 1$
\begin{eqnarray}
H_{q} &=&-\bar{\epsilon} ^{q}N
-G_{q}F_{0}G_{0}-F_{q}(F_{+1}G_{+1}+F_{-1}G_{-1})
-\frac{1}{2}E_{q}(\{ T_{+},T_{-}\} -\left[ \frac{N}{2\Omega }\right])
\nonumber \\
&&-C_{q}2\Omega \left[ \frac{1}{\Omega }
\right] (\left[ K_{0}\right] ^2_{\frac{1}{2\Omega }}
- \left[ \Omega \right]^2 _{\frac{1}{2\Omega }}) -D_{q}\Omega \left[
\frac{1}{\Omega }\right] \left[ T_{0}\right] ^2_{\frac{1}{2\Omega }},
\label{qH}
\end{eqnarray}
where $\epsilon ^{q}=\bar{\epsilon} ^{q}+(\frac{1}{2}-2\Omega
)C_{q}+\frac{D_{q}}{4}>0$ is the Fermi level of the nuclear system,
$K_{0}$ is related
to $N$ (Table 1), $G_{q},\ F_{q},\ E_{q},\ C_{q}$ and $D_{q}$ are
constant interaction
strength parameters and in general they may be different than the non-deformed
phenomenological parameters.
\newline

{\bf Matrix elements of the Hamiltonian --} In the $SU^{0}(2)$
limit ($pn$-coupling) the energy
eigenvalue of the non-deformed pairing interaction $GA_{0}B_{0}$ is
\begin{eqnarray}
\hspace{-0.5in}
\varepsilon _{pn}=\frac{G}{\Omega }n_{0}\frac{2\Omega -n+n_{0}+1}{2}
=\frac{G}{8\Omega }(n-2\nu _{0})(4\Omega -n-2\nu _{0}+2)  \label{en0}
\end{eqnarray}
and in the $SU^{\pm }(2)$ limit (like-particle coupling) the
energy of the non-deformed
pairing interaction $FA_{\pm 1}B_{\pm 1}$ is
\begin{eqnarray}
\hspace{-0.5in}
\varepsilon _{pp(nn)} &=&\frac{F}{\Omega }n_{\pm 1}(\Omega +n_{\pm 1}-N_{\pm
}+1) =\frac{F}{4\Omega }\left( N_{\pm }-\nu _{1}\right) (2\Omega -N_{\pm }-\nu
_{1}+2).  \label{en_ppnn}
\end{eqnarray}
In each limit, $\nu _{0}=n_{1}+n_{-1}$ and $\nu _{1}={n}_{0}$ are the
respective seniority quantum numbers that count the number of remaining
pairs that can be formed after coupling the fermions in the primary pairing
mode and they vary by $\Delta \nu _{0,1}=2$.

To investigate the influence of the deformation on the pairing interaction,
the eigenvalue of the deformed pairing Hamiltonian is expanded in orders of $
\varkappa $ ($q=e^{\varkappa }$) in each limit
\begin{eqnarray}
&\varepsilon _{pn}^{q}&=G_{q}\left[ \frac{1}{2\Omega }\right] \left[ \frac{
n-2\nu _{0}}{2}\right] _{\frac{1}{2\Omega }}\left[ \frac{4\Omega -n-2\nu
_{0}+2}{2}\right] _{\frac{1}{2\Omega }}  \nonumber\\
=\frac{G_{q}}{G}&\varepsilon _{pn}&\{1+\varkappa ^{2}\frac{
(n_{0}^{2}-4\Omega ^{2}-1)+\left( \frac{2\Omega \varepsilon
_{pn}}{n_{0} G}\right) ^{2}}{24\Omega ^{2}}+O(\varkappa ^{4})\},
\label{qHexpand0}
\end{eqnarray}

\begin{eqnarray}
&\varepsilon _{pp(nn)}^{q}& =F_{q}\rho _{\pm }\left[ \frac{1}{\Omega }
\right] \left[ \frac{N_{\pm }-\nu _{1}}{2}\right] _{\frac{1}{\Omega }}\left[
\frac{2\Omega -N_{\pm }-\nu _{1}+2}{2}\right] _{\frac{1}{\Omega }}
\nonumber \\
=\frac{F_{q}}{F}&\varepsilon _{pp(nn)}&\{1\pm \varkappa 
\textstyle{\frac{1+2\Omega }{
4\Omega }}+\varkappa ^{2}\frac{(n_{\pm 1}^{2}+\frac{\Omega ^{2}}{2}-\frac{5}{8
})+\left( \frac{\Omega \varepsilon _{pp(nn)}}{n_{\pm 1} F}\right) ^{2}}{
6\Omega ^{2}}+O(\varkappa ^{3})\},
\label{qHexpand_pm}
\end{eqnarray}
where the non-deformed energies (\ref{en0}) and (\ref{en_ppnn}) are the
zeroth order approximation of the corresponding deformed pairing energies.
While the proton-neutron interaction is even with respect to the deformation
parameter $\varkappa ,$ the identical particle pairing includes odd
terms as well through the coefficient $\rho _{\pm }.$ The
expansions in the pairing limits ((\ref{qHexpand0}) and (\ref{qHexpand_pm})) introduce
non-linear terms with respect to the pair numbers, space dimension and the non-deformed
pairing energies. They serve as a simple example of the contribution of the
$q$-deformation compared to the non-deformed model, which is a straightforward result
of the quantum definition.

In general, the Hamiltonian (\ref{clH}) is not diagonal in the basis set
(Table 2b). The linear combinations of the basis states describe the
spectrum of the isovector-paired $0^{+}$ states for a given nucleus.
The pairing
Hamiltonian $H_{pair}$ ((\ref{clH}) with $E=C=D=0$ and $\epsilon =0$)\
gives a transition between the states with different kinds of pairing while
preserving the total number of pairs, $N$, that is, two $pn$
pairs scatter into a $pp$ and a $nn$ pair, and vice versa
\begin{eqnarray}
\hspace{-1in}
\left| H_{pair}\right| \left| n_{1},n_{0},n_{-1}\right) &=&
(\varepsilon
_{pn}+\varepsilon _{pp}+\varepsilon _{nn})\left| n_{1},n_{0},n_{-1}\right)
-\frac{G}{\Omega }n_{1}n_{-1}\left| n_{1}-1,n_{0}+2,n_{-1}-1\right)
\nonumber \\
&&-\frac{F}{\Omega }n_{0}(n_{0}-1)\left| n_{1}+1,n_{0}-2,n_{-1}+1\right) ,
\label{HpairME}
\end{eqnarray}
where $\varepsilon _{pn,pp,nn}$ are given in (\ref{en0}) and
(\ref{en_ppnn}) and $n_{1},n_{0},n_{-1}$ are particle or hole pairs.

We are also able to find an analytical form of the $q$-deformed analog of
(\ref{HpairME})
\begin{eqnarray}
\hspace{-1in}
\left| H_{q,pair}\right| \left| n_{1},n_{0},n_{-1}\right) &=&
(\varepsilon _{pn}^{q}+\varepsilon _{pp}^{q}+\varepsilon _{nn}^{q})\left|
n_{1},n_{0},n_{-1}\right)-\frac{G}{\Omega }\tilde{n}_{1}\tilde{n}_{-1}\left|
n_{1}-1,n_{0}+2,n_{-1}-1\right)   \nonumber \\
&&-\frac{F}{\Omega }\frac{\sqrt{\rho _+ \rho _-}}{\left[
2\right]}\sum_{k=1}^{n_{0}-1}
S_q(k)
\left| n_{1}+1,n_{0}-2,n_{-1}+1\right) ,  \label{qHpairME}
\end{eqnarray}
where  $\varepsilon _{pn,pp,nn}^{q}$
are given in (\ref{qHexpand0}) and (\ref{qHexpand_pm}). We define
$\tilde{n}_{\pm
1}\equiv \frac{1}{[2]}\sqrt{\rho_+ \rho_-}\left[ n_{\pm 1}\right]
_{\frac{1}{2\Omega
}}\left[ 2_{n_{\pm 1}-\Omega -1/2}\right] _{\frac{1}{2\Omega }}$
$\stackrel{q\rightarrow
1}{\rightarrow }n_{\pm 1}$, $S_q(k)\equiv
[2_{k-\Omega -1/2}]_{\frac{1}{2\Omega }} \sum_{i=0}^{k-1}\frac{\left[ 2\right]
^{i}}{2^{i}}[2_{k-1-i}]_{\frac{1}{2\Omega }}\stackrel{q\rightarrow
1}{\rightarrow }4k$ and
$[2_{X}]_{
\frac{1}{2\Omega }}\equiv \frac{\left[ 2X\right] _{\frac{1}{2\Omega }}}{
\left[ X\right] _{\frac{1}{2\Omega }}}\stackrel{q\rightarrow
1}{\rightarrow }2$.

\section{Applications to nuclear structures}

\subsection{ 0$^{+}$-state energy for even-{\it A} nuclei. Discussion
of the results.}

The eigenvalues of the Hamiltonians (\ref{clH}) and (\ref{qH}) describe
nuclear isovector-paired $0^{+}$ state energies, which are fit to experimental
values \cite{AudiWapstra,Firestone}. For even-even nuclei and
for some odd-odd nuclei ($Z\approx N$), the lowest
$0^{+}$ state is the nuclear ground state and the positive value of
its energy is defined as the
binding energy, $|BE|$. The binding energy of a nucleus is an important
quantity because it is
related to the nuclear mass and lifetime. Other odd-odd nuclei have a
higher-lying $0^{+}$ excited state which is an isobaric analog of the
corresponding even-even neighbors.

The phenomenological parameters in (\ref{clH}) and (\ref{qH}) are determined
by a non-linear least-squares fit of the lowest isovector-paired $0^{+}$ state
energies (maximum eigenvalues of $|H|$ (\ref{clH}) and $|H_q|$
(\ref{qH})) to the Coulomb
corrected experimental values:
\begin{equation}
E_{0}^{\exp }(N_{+},N_{-})=|E_{\exp }(N_{+},N_{-})|-|E_{\exp
}|_{core}+V_{Coul}(N_{+},N_{-}),  \label{EexpCoul}
\end{equation}
where the binding energy of the core $|E_{\exp }|_{core}$ is subtracted in
order to focus only on the contribution from the valence shell. The
energies need to be corrected for the Coulomb repulsion since it is not
accounted for by the model Hamiltonian. In (\ref{EexpCoul}) the
Coulomb potential
is taken relative to the core
and is derived in
\cite{RetamosaCaurier}.

The parameters and statistics, obtained from the fitting procedure, are
shown in Table 3. In both the non-deformed (``non-def" column) and deformed cases
(``$q$-def" column),
three groups of even-$A$ nuclei are considered: $({\bf I})$ $1d_{3/2}$ $(\Omega =2)$
with a core $_{16}^{32}S$ (Table 2a); $({\bf II})$ $
1f_{7/2}$ $(\Omega =4)$ with a core $_{20}^{40}Ca$ (Table 2b); and ({\bf III})
$1f_{5/2}2p_{1/2}2p_{3/2}1g_{9/2}$ major shell ($\Omega =11$) with a core $_{28}^{56}Ni$.
In each group, the number of the valence protons (neutrons) varies in the range $N_{\pm
}=0,\ldots ,2\Omega
$ and the total number of nuclei that enter into the $Sp(4)$ systematics is $2\Omega (\Omega
+1)+1$ (13 for $({\bf I})$, 41 for $({\bf II})$ and 265 for $({\bf III})$). The residual
sum of squares
$ S\equiv
\left(
\left| E_{0}^{th}\right|-\left| E_{0}^{\exp }\right| \right) ^{2}$ and the chi-statistics
$\chi
\equiv \sqrt{\frac{S}{N_{d}-n_{p}}}$ define the goodness of the fit, where $n_{p}$ is the
number of the fitting parameters and
$N_{d}$ is the number of nuclei with available data
($N_d$ is
$13$ in ({\bf I}), $36$ in ({\bf II}) and $100$ in ({\bf III})).

Analysis of the results (Table 3) shows that for $1d_{3/2}$ the pairing parameters
are almost equal $(G\approx F)$ as it is expected for light nuclei, and they differ, $G>F,$
for
$1f_{7/2}$ by $0.07$ and for $({\bf III})$ by $0.06$.
Based on the estimation of
the parameters (Table 3) and the correlations (\ref{eta_FG}) the extent to
which the symmetry in each limit is broken can be evaluated. In the limit $
SU^{\tau }(2)$, the breaking of the isospin invariance $\eta _{2}/\eta _{1}$
is in general small for light nuclei ($\eta _{2}/\eta
_{1}=0.090$ for $1d_{3/2}$ and $\eta _{2}/\eta _{1}=0.\,133$ for 
$1f_{7/2}$), which is in agreement with the
experimental data for this region. For medium nuclei in the
$1f_{5/2}2p_{1/2}2p_{3/2}1g_{9/2}$ major shell the isospin breaking is
significantly greater, $\eta _{2}/\eta _{1}=0.\,628$.

\smallskip \noindent {\bf Table 3.} Fit parameters and statistics. 
$G$, $F$, $C$, $D$, $\epsilon $ and $\chi $ are in$\ MeV,\ S\ $is in$\ MeV^{2}$.
Quantities marked with the symbol * are fixed for a given fit. 
\[
\begin{tabular}{|c||c|c||c|c||c|c|}
\hline
& \multicolumn{2}{c||}{ ({\bf I}) } & \multicolumn{2}{c||}{({\bf II})} &
\multicolumn{2}{c|}{({\bf III})} \\ \hline 
& \multicolumn{2}{c||}{ $1d_{3/2}$ } & \multicolumn{2}{c||}{ $1f_{7/2}$ } &
\multicolumn{2}{c|}{ $1f_{5/2}2p_{1/2}2p_{3/2}1g_{9/2}$ } \\ \hline 
& non-def & $q$-def & non-def & $q$-def & $\ $non-def$\ $ & $q$-def\\
\hline \hline
$\varkappa $ & {\small 0*} & {\small -0.015} & {\small 0*} & {\small 0.124} & {\small 0*} &
{\small 0.215} 
\\ \hline \hline
$q=e^{\varkappa }$ & {\small 1*} & {\small 0.985} & {\small 1*} &
{\small 1.132} & {\small 1*} & {\small 1.240} \\ \hline
$G/{\Omega }$ & {\small 0.709} & {\small 0.709*} & {\small 0.525} &
{\small 0.525*} & {\small 0.352} & {\small 0.352*} \\ \hline
$F/{\Omega }$ & {\small 0.702} & {\small 0.702*} & {\small 0.453} &
{\small 0.453*} & {\small 0.296} & {\small 0.296*} \\ \hline
$C$  & {\small 0.815} & {\small 0.815*} & {\small 0.473} & {\small 0.473*} & {\small 0.190}
& {\small 0.190*} \\
\hline
$D$ & {\small -1.282} & {\small -1.282*} & {\small -0.971} & {\small -0.971*} & {\small
-0.796} & {\small -0.796*} \\ \hline
$E/{(2\Omega )}$ & {\small -1.409} & {\small -1.409*} & {\small -1.120}
& {\small -1.120*}  & {\small -0.489} & {\small -0.489*} \\ \hline
$\epsilon $ & {\small 9.012} & {\small 9.012*} & {\small 9.359} &
{\small 9.359*}  & {\small 9.567} & {\small 9.567*}  \\ \hline \hline
$S$ & {\small 1.720} & {\small 1.719} & {\small 16.095} & {\small 15.673} & {\small
300.284} & {\small 238.280} \\ \hline
$\chi $ & {\small 0.496} & {\small 0.378} & {\small 0.732} & {\small 0.669} & {\small1.787}
& {\small 1.551} \\ \hline
\end{tabular}
\label{tab:fitStat}
\]

\begin{figure}[th]
\centerline{\hbox{\epsfig{figure=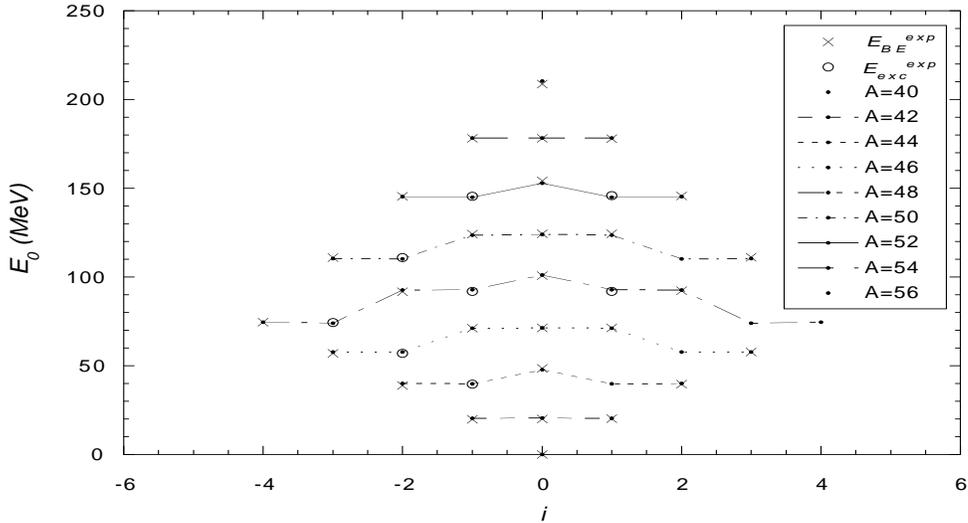,width=13cm,height=7cm}}}
\caption{ Coulomb corrected $0^{+}$ state energy $E_{0}$ vs. the isospin
projection
$i$ for  the isotopes of
nuclei with $Z=20$ to $Z=28$ in the $1f_{7/2}$ level$,$ $\Omega _{7/2}=4.$ The
experimental binding energies $E_{BE}^{\exp }$ (symbol ``$\times $'') are
distinguished from the experimental energies of the isobaric
analog $0^{+}$
excited states
$E_{exc}^{\exp }$ (symbol ``$\circ $''). Each line connects theoretically
predicted energies of an isobar sequence. The nuclei for which experimental
data is not available are represented only by their predicted $0^{+}$ state
energy. }
\label{BindEn}
\end{figure}

An observation about the $pn$-pairing strength is that most of
$pn$-coupling study has been done assuming
good isospin, that is $F=G$.
However, the free nucleon-nucleon data \cite{Lawson} 
indicates that the $\tau =1\
pn$-pairing strength ($G$) is slightly bigger than the like-particle pairing
strength ($F$), which is confirmed by the fits presented in Table 3 and is
investigated in other studies  \cite{CivitareseReboiroVogel,GoswamiChen}.
There are many different values for the like-particle pairing strength
used in  literature. The most common
value is taken to be proportional to $1/A$
\cite {Lane,KermanLawsonMacfarlane,KisslingerSorensen,BohrM69,DudekMS} 
and is consistent with the experimental pairing gaps derived from the
odd-even mass differences
\cite {NilssonP61,RingS80}. The values of
$F$, obtained by our theoretical symplectic model, fall within the 
limits of their estimation. In this way, they are expected to reproduce the low-lying
vibrational spectra of spherical nuclei in the
$SU^{\pm }(2)$ limit. When the results from all the three non-deformed fits
are considered (Table 3), the identical-particle parameter $F/\Omega $ is found to decrease
with the mass number as $23.\,9/A$. In a similar way, one can find the dependence of the
$pn$ pairing strength parameter $G/\Omega $ on the mass number to be $25.\,7/A$.

The estimate of the parameters (Table 3) reveals the properties of the
nuclear interaction as interpreted by connection (\ref{V_FG}). The $J=0$
pairing interaction ($V_{P_0}$) is always attractive, while the
overall high-$J$ component of identical-nucleon coupling
$\left\langle \pm \pm \left| V\right| \pm \pm \right\rangle$ might be
repulsive. The $J>0$ proton-neutron ``direct'' interaction $\left\langle
-+\left| V\right| -+\right\rangle $ is attractive, but not the ``exchange'' part of it
$\left\langle -+\left| V\right| +-\right\rangle $ ($E<0$).

In all cases there is a good agreement with experiment (small $\chi $),
as can be seen in Table 3{\bf ,} as well as in Figure \ref{BindEn} for
region $({\bf II}).$ Part of our results, namely for the  binding energies (but not
for the excited $0^+$ state energies), can be compared to other theories. A direct
comparison of the chi statistics is impossible because of the different data sets 
and energy levels determined by the various theories. However, if we select only the 
data subsets that are equivalent for the nuclei in the
$1d_{3/2}$ and/or $1f_{7/2}$, our results are much closer to the 
experimental numbers than those for the Hartree-Fock-Bogoliubov ($HFB$) model
\cite{SandhuRustgi} and the semi-empirical model \cite{Moler} and comparable with those of
the $jj$-coupling shell model \cite{TalmiSimpleModels,TalmiThieberger}
and the isovector and isoscalar pairing plus quadrupole model \cite 
{HasegawaKanekoPRC59}.
In this way the simple $Sp(4)$ model is tested and proves its validity when
applied to light nuclei in single-$j$ level. In this region many symmetries are
conserved allowing for a possible reduction of the number of fitting parameters. However,
the free parameters in the fits presented in Table 3 reflect the symmetries observed
in light nuclei and the non-negligible symmetry breaking in medium-mass
nuclei.
\newline

\begin{figure}[th]
\centerline{\hbox{\epsfig{figure=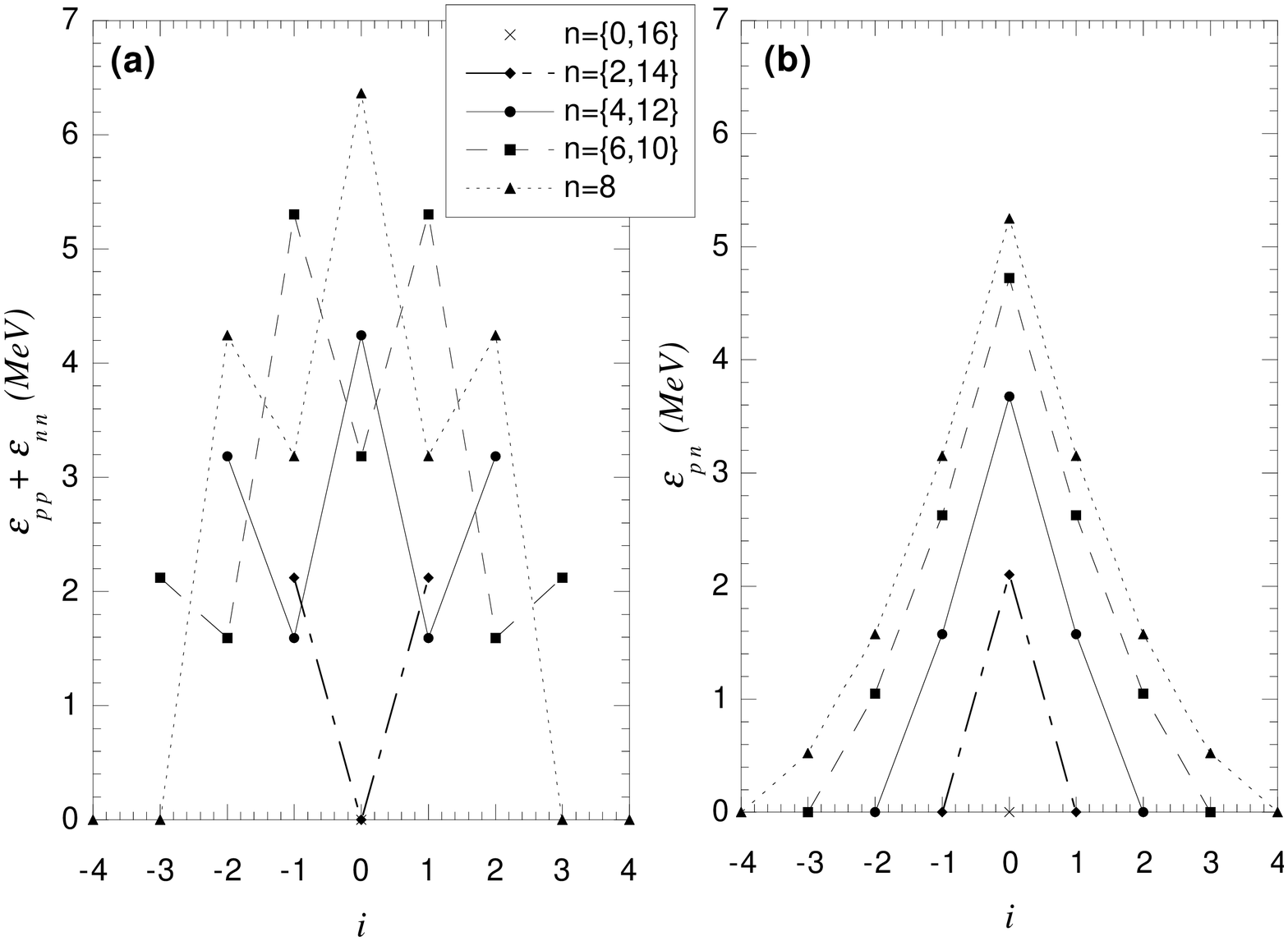,width=14.5cm,height=6.0cm}}}
\caption{Identical ($G=0,\ F=1.81$ (a)) and non-identical ($G=2.10,\ F=0$
(b)) particle pairing energies vs. $i$ for isobars with $A=40$ to $A=56$ in the
$1f_{7/2}$ level}
\label{Ca_pn_ppnn}
\end{figure}

{\bf Properties of the pairing interaction --} A model with the
$Sp(4)$ dynamical symmetry permits an independent investigation of the different kinds
of pairing interactions in
the limiting cases of the non-deformed (\ref{en0}), (\ref{en_ppnn}) as well as
the deformed versions (\ref{qHexpand0}), (\ref{qHexpand_pm}) of the theory. In
the $SU^{\pm }(2)$ limit, the symplectic model reproduces the
properties of the identical-nucleon pairing ($\varepsilon _{pp}+\varepsilon
_{nn}$) (\ref{en_ppnn}), for which the usual parabolic dependence
of $ \varepsilon _{pp(nn)}$ on $N_{\pm }$\ holds
\cite{KermanLawsonMacfarlane,Kerman,TalmiSimpleModels,HeydeGreiner}.
The dependence of the like-particle
energy on the isospin projection
$ i $ (Figure \ref{Ca_pn_ppnn}(a)) reveals another property of the pairing
mode, a $\triangle i=1$ staggering of the identical-nucleons pairing
energies of the odd-odd and even-even nuclei.
In contrast with this, the $pn$ limit ($\varepsilon _{pn}$) shows a smooth behavior (Figure
\ref{Ca_pn_ppnn} (b)). The $SU^{0}(2)$ limiting case yields a
proton-neutron coupling
that has its maximum when $N_{+}=N_{-}$ ($i=0$)$,$ which is consistent with
$\alpha -$clustering theories \cite{HasegawaKanekoGRS,RopkeSS} and the
charge independence in the region of light nuclei when protons and
neutrons fill the same shell \cite{MacchiaveliPLB,Vogel}. In both limits ($SU^{\pm }(2)$ and
$SU^{0}(2)$ ), the pairing energy decreases when the difference between proton and
neutron numbers increases.
\begin{figure}[b]
\centerline{\hbox{\epsfig{figure=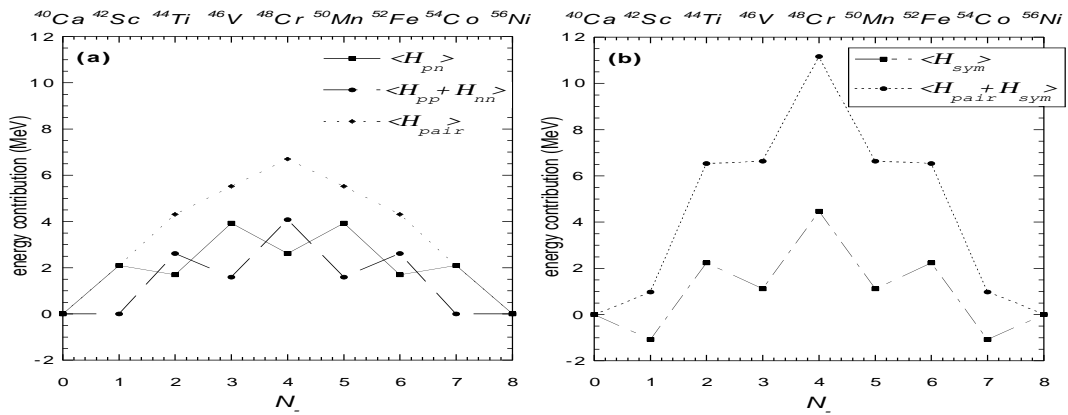,width=14.5cm,height=6.5cm}}}
\caption{Lowest isovector-paired $0^{+}$ state energies vs. $N_{-}$ when $
N_{+}=N_{-}$ for the nuclei in $1f_{7/2}$: (a) pairing energies: $pn,$ $pp+nn$
and total; (b) symmetry energy ($E$ term in (\ref{clH})) and
total pairing +
symmetry energy}
\label{CaNeqZ}
\end{figure}

In most nuclei the different pairing interactions coexist and 
the contribution of each of the pairing modes,
$\left\langle H_{pp}+H_{nn}\right\rangle $ $=$ $ F\left\langle
A_{+1}B_{+1}+A_{-1}B_{-1}\right\rangle $ and $\left\langle H_{pn}\right\rangle
=G\left\langle A_{0}B_{0}\right\rangle $, to the total pairing energy
$\left\langle H_{pair}\right\rangle =\left\langle H_{pn}\right\rangle
+\left\langle H_{pp}+H_{nn}\right\rangle $ can be investigated. A $\triangle
N=2$ staggering exists for both pairing interactions (Figure \ref{CaNeqZ}(a))
\cite{EngelLangankeVogel,Langanke97,KanekoHZPRC59}.
For $N_{+}=N_{-}$ odd-odd nuclei the $\tau =1\ pn$ pairs give the dominant
contribution, while for the even-even $N_{+}=N_{-}$ nuclei both
pairing modes
contribute almost equally with a slightly greater like-particle contribution.
Although difference between even-even and odd-odd nuclei exists in each pairing
contribution, the total pairing energy has a surprisingly smooth behavior.
The contribution from the symmetry term ($E$ term in (\ref{clH})) \cite
{ZeldesLirin76,Vogel} restores the staggering, as it
decreases the energy of the odd-odd nuclei with respect to
their even-even neighbors  (Figure
\ref{CaNeqZ}(b)).

\begin{figure}[th]
\centerline{\hbox{\epsfig{figure=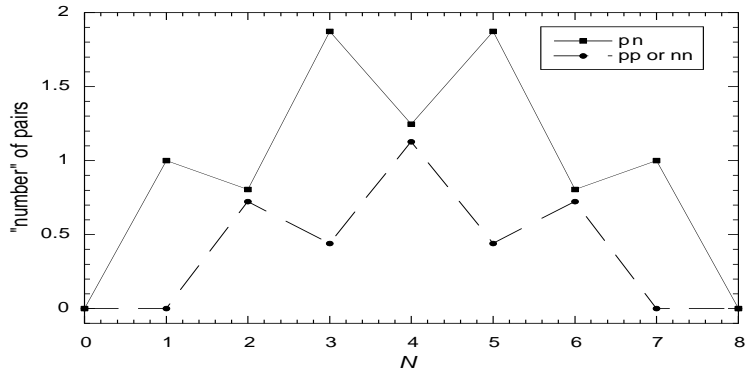,width=10cm,height=5cm}}}
\caption{Non-identical ($pn$) and identical ($pp$ or $nn$) pairing
``numbers'' vs. $N_{-}$ for the $N_{+}=N_{-}$ nuclei with $Z=20$ to $Z=28$
in $1f_{7/2}$}
\label{CaNpairs}
\end{figure}

Rough measures for the number of $pn$ and like-particle pairs are the
quantities $\frac{1}{G}\left\langle H_{pn}\right\rangle $ and $\frac{1}{F}
\left\langle H_{pp}+H_{nn}\right\rangle $, respectively, which are related
to the pairing gaps \cite{EngelLangankeVogel,Dobes}. The ``number''
of $pn$ pairs (Figure \ref{CaNpairs}) is bigger than the ``number'' of $
pp(nn)$ pairs for odd-odd $N_{+}=N_{-}$ nuclei, and is of the same order as for
the even-even nuclei \cite{EngelLangankeVogel,Langanke97,KanekoHZPRC59}.
\newline

{\bf Deformed non-linear model --} To investigate the role of the $q$-deformation, we
performed again the fitting procedures for the same regions (${\bf (I)}$,
${\bf (II)}$ and ${\bf (III)}$) but using the deformed Hamiltonian (\ref{qH}).
For each group of nuclei, the outcome  of a fit with all possible
parameters ($G_{q},\ F_{q},\ E_{q},\ C_{q},\ D_{q},\ \epsilon ^{q}$ and $q$) indicates that
the introduction of the $q$-deformation does not vary the rest of the parameters.
Based on  this result, we considered the deformation to be independent of the other
parameters and varied only $q$ in the fit (the rest of the parameters were 
kept fixed with values obtained from the non-deformed fit). The
results are shown in the ``$q$-def" columns in Table 3. The fits with and without a
deformation can be compared by using the residual sum of  squares ($S$), which is always
smaller in the deformed case (Table 3).

Although it stands in contrast with other $q$-deformed applications
\cite{Sharma,SharmaSh}, the decoupling of the $q$-deformation from the interaction strengths
is not an assumption but results from comparisons to experimental data over total of $149$
nuclei.  It implies that while leaving the strength of the
two-body interactions unchanged, the $q$-deformation allows one to take into account, in a
prescribed way, complicated dependence of the energy eigenvalues on the
number of nucleons/pairs and space dimension that cannot be reproduced by any two-body
interaction (for example, see (\ref{qHexpand0}) and (\ref{qHexpand_pm})). Moreover, similar
terms are expected to arise from higher-order interactions between the particles.
In this way the $q$-parameter introduces some non-linear
residual interaction not present in the two-body Hamiltonian (\ref{clH}).

\begin{figure}[th]
\centerline{\hbox{\epsfig{figure=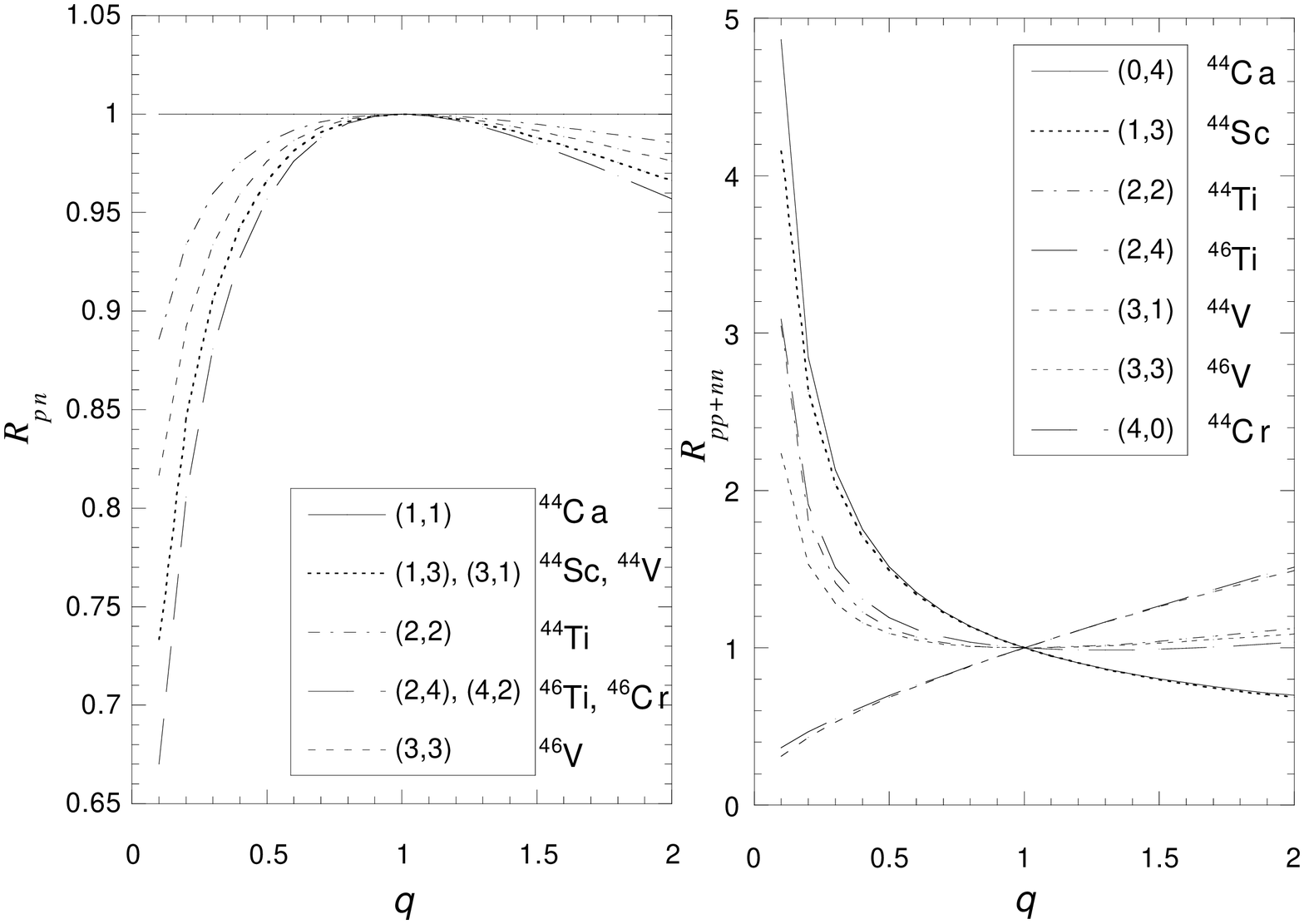,width=14.5cm,height=7.0cm}}}
\caption{Ratios $R_{pn}$ and $R_{pp+nn}$ vs. $q$ for several nuclei with a
typical behavior in the $1f_{7/2}$ level}
\label{qCa_F_G}
\end{figure}

The observed independence of the pairing strengths on the $q$-parameter,
suggests that while the deformation does not change the strength to couple two
particles, it can model many-pair effects and can influence the
energy spectrum. As an illustration, in each of the dynamical limits we investigate the
quantities
$R_{pn}=\varepsilon _{pn}^{q}/\varepsilon _{pn} $ and
$R_{pp+nn}=(\varepsilon _{pp}^{q}+\varepsilon _{nn}^{q})/(\varepsilon
_{pp}+\varepsilon _{nn})$ that give an additional
contribution to the pairing energy in the deformed case (Figure \ref{qCa_F_G})
(compare to the analytical expansion with respect to $\varkappa $ of the energies,
(\ref{qHexpand0}) and (\ref{qHexpand_pm})). In the limit of
$pn$-pairing,
$R_{pn}$ does not significantly change when $q$ is close to one and it decreases for
all $q \neq 1$. The ratio
$R_{pp+nn}$ increases (decreases) monotonically with $q$ only for nuclei
with a primary $pp$ ($nn$) coupling.
Even though both $SU_{q}^{\pm }(2)$ groups are complementary, the
different behavior
of the multiplication constants $\rho _{\pm }$ (Table 1) is responsible for
different impact of the deformation in various isotopes. This
accounts for the
differences in the experimental data between mirror nuclei even after the
Coulomb energy correction. In the limit of identical-particle
coupling, when $q$
increases from one ($q>1$) neutron pairs are less bound and proton pairs give a
larger pairing gap, and vice  versa for $q<1$. In this way, the deformation
parameter can determine the degree to which the $pp$ coupling differs
from the $nn$ coupling. 

The significance of the higher-order terms that enter through the $q$-deformed
theory can be estimated through a comparison with experiment. In general, the fitting 
procedures determine values for $\varkappa $ (Table 3) that are small. The reason may
be  that while higher-order effects may be significant in nuclei they probably cancel on
average when the $q$-parameter is one and the same for all nuclei. However, in two of
the cases, $({\bf II})$ and $({\bf III})$, it is of an order of magnitude greater than the
estimation of other physical applications (\cite{Bonatsos} and references there) and for
the shell
$1f_{7/2}$, our value ($q=1.\,132$) is comparable to the values obtained in a $q$-deformed
like-particle seniority model \cite{Sharma}: $q=1.\,1585$ for the
neutron pairs and $ q=1.\,1924$ for protons. 
For the nuclei in the multi-$j$ shell our model yields a bigger $q$-parameter than
for the lighter nuclei
in single-$j$ shell (Table 3), where the small
number of valence nucleons is not sufficient to build strong non-linear correlations. This
suggests that the
$q$-deformation is more significant for masses $A>56$. 

\subsection{Predicted energies}

The fitting procedure not only estimates the magnitude of the pairing
strength and describes the type of the dominant coupling mode, it also
can be used to predict nuclear energies that have not been measured. From
the fit for the $1f_{7/2}$ case the binding energy of the proton-rich
$^{48}Ni$ nucleus is estimated to be $348.19$ MeV, which is by $0.07\%$ greater than
the sophisticated semi-empirical estimate of \cite{Moler}. Likewise, for the
odd-odd nuclei that do not have measured energy spectra the theory 
can predict the energy of
their lowest $0^{+}$ isobaric analog state: $358.75$ MeV ($^{44}V$), 
$359.49$ MeV ($
^{46}Mn$), $357.56$ MeV ($^{48}Co$), $394.16$ MeV ($^{50}Co$). The 
predicted energies
are calculated for $q=1.132$ (Table 3 ({\bf II})) as the fit with 
deformation has a
smaller uncertainty compared to the non-deformed one. The $Sp_{(q)}$ model predicts
the relevant $0^+$ state energies for additional 165 even-$A$ nuclei in the medium mass
region ({\bf III}). The binding energies for 25 of them are also calculated in \cite{Moler}.
For these even-even nuclei,
we predict binding energies that on
average are by $0.05\%$ (non-deformed case) and by $0.008\%$ (for
$q=1.240$) less than the semi-empirical approximation \cite{Moler}.

\section{Conclusion}

We constructed a model with a symplectic dynamical symmetry group
$Sp_{(q)}(4)\supset U(1)\otimes SU_{(q)}(2)$ in the non-deformed limit as well as in
the $q$-deformed generalization. A phenomenological Hamiltonian was
written in terms of the generators of the group and this in turn was used to describe
pairing correlations in nuclei. The relation of this approach to a general microscopic
pairing Hamiltonian was
obtained. The theory was tested by fitting calculated energies to
the relevant experimental
$0^{+}$ state energies for single-$j$ levels, namely 
$1d_{3/2}$ and $1f_{7/2}$, and for a multi-$j$ $1f_{5/2}2p_{1/2}2p_{3/2}1g_{9/2}$ shell.
In general, the fitting procedure yielded results that were  in good agreement with the
experiment. The theory predicted the lowest $0^{+}$ isovector-paired 
state energy of nuclei with a deviation of at most $ 0.5\%$ in the energy range considered. 
It was used to predict the binding  energy for even-even nuclei and the lowest 
isovector-paired $0^{+}$ state energy of odd-odd nuclei in the proton-rich region. The 
phenomenological pairing parameters and
the strength of the pairing interaction were determined. In agreement with
experimental analysis, breaking of the isospin invariance, $\eta _2/\eta _1 \neq 0$, and
isospin mixing, $G>F$, is observed, except for light nuclei in the $1d_{3/2}$ level.

The theoretical model with $Sp(4)$ dynamical symmetry and its
$q$-deformed version was used to investigate in greater detail
the properties of
the isovector pairing interaction. The study reveals that the pairing energy
along with the symmetry energy are responsible for the experimentally observed
staggering between even-even and odd-odd nuclear isovector-paired $0^{+}$ state
energies. Overall, the results show that the symplectic model can be
used  to provide
a reasonable description of isovector-paired $0^{+}$ states in
nuclei, confirming in its limit
results previously published for like-particle pairing correlations.
When only the
nuclei with $0^+$ ground states are considered, the non-deformed
$Sp(4)$ model is
comparable, for the region of light nuclei, to earlier theories. At
the same time it
gives some insight into the study of symmetry breaking  
and isovector pairing
correlations, and it is  based on a simple
approach that is applicable in a broad region of the nuclear chart,
including odd-odd and exotic nuclei.

The $q$-deformed case gives the best overall results. The $Sp_q(4)$ dynamical symmetry
approach yielded a $q$-deformed exact solution and we were able to derive the
$q$-deformed matrix elements of the interaction in a simple analytical form. In addition
to the broken symmetries of the non-deformed model, the $q$-deformation breaks the
symmetry between protons and neutrons, which again is small for light nuclei and
consistent with experiment. The introduction of $q$ leads to 
a decrease of the like-particle pairing gap for neutrons and an increase for 
protons as $q$ increases from one. The $q$-parameter was found decoupled from the
interaction strength parameters. This observation suggests that while the deformation does
not influence the two-body interaction, it introduces higher-order interactions between the
particles, which are neglected in most non-deformed models. The $q$-deformation is
mass and shell dimension dependent  and its effects are more significant in the medium mass
region. In the present study, the $q$-parameter was found as high as $1.240$ and is
expected to be greater if its influence is not averaged over all nuclei in a major shell.
This suggests the need for a more
elaborate investigation of the role of the
$q$-deformation in each individual nucleus and the relation of the $q$-parameter to the
underlying nuclear structure.

\vskip .5cm This work was supported by the US National Science 
Foundation, Grant
Numbers 9970769 and 0140300. The authors appreciate the encouraging 
discussions of this work with
Professor Feng Pan, Dr. Chairul Bahri and Dr. Carl E. Svensson, as 
well as with Dr. Vesselin G. Gueorguiev,
whom we thank also for his computational {\small MATHEMATICA}\ programs for
non-commutative algebras.

\end{document}